\documentclass[aip,jcp,reprint]{revtex4-2}
\usepackage[utf8]{inputenc} 
\usepackage[T1]{fontenc}       
\usepackage{amssymb}
\usepackage{amsmath}
\usepackage{comment} 

\usepackage{float} 
\usepackage{graphicx} 
\usepackage{color}
\usepackage[singlelinecheck=false,justification=raggedright]{caption}
\usepackage[parfill]{parskip} 

\usepackage{algorithm}
\usepackage{algpseudocode}

\usepackage{booktabs} 
\usepackage{makecell}

\newcommand{\mean}[1]{\langle #1 \rangle}

\makeatletter
\newcommand*{\rom}[1]{\expandafter\@slowromancap\romannumeral #1@}
\makeatother

\begin{document}
\title{Convergence Analysis of the Stochastic Resolution of Identity: Comparing Hutchinson to Hutch++ for the Second-Order Green's Function}

\author{Leopoldo Mej\'ia}
\email{leopoldo.mejia@berkeley.edu}
\affiliation{Department of Chemistry, University of California, Berkeley, California 94720, USA}
\affiliation{Materials Sciences Division, Lawrence Berkeley National Laboratory, Berkeley, California 94720, USA}

\author{Sandeep Sharma}
\email{sandeep.sharma@colorado.edu}
\affiliation{Department of Chemistry, University of Colorado Boulder, Boulder, Colorado 80309, USA}

\author{Roi Baer}
\email{roi.baer@huji.ac.il}
\affiliation{Fritz Haber Center for Molecular Dynamics, Institute of Chemistry, The Hebrew University of Jerusalem, Jerusalem 91904, Israel}

\author{Garnet Kin-Lic Chan}
\email{garnetc@caltech.edu}
\affiliation{Division of Chemistry and Chemical Engineering, California Institute of Technology,
Pasadena, California 91125, USA}

\author{Eran Rabani}
\email{eran.rabani@berkeley.edu}
\affiliation{Department of Chemistry, University of California, Berkeley, California 94720, USA}
\affiliation{Materials Sciences Division, Lawrence Berkeley National Laboratory, Berkeley, California 94720, USA}
\affiliation{The Raymond and Beverly Sackler Center of Computational Molecular and Materials Science, Tel Aviv University, Tel Aviv 69978, Israel}

\begin{abstract}
Stochastic orbital techniques offer reduced computational scaling and memory requirements to describe ground and excited states at the cost of introducing controlled statistical errors. Such techniques often rely on two basic operations, stochastic trace estimation and stochastic resolution of identity, both of which lead to statistical errors that scale with the number of stochastic realizations ($N_{\xi}$) as $\sqrt{N_{\xi}^{-1}}$.  Reducing the statistical errors without significantly increasing $N_{\xi}$ has been challenging and is central to the development of efficient and accurate stochastic algorithms.  In this work, we build upon recent progress made to improve stochastic trace estimation based on the ubiquitous Hutchinson's algorithm and propose a two-step approach for the stochastic resolution of identity, in the spirit of the Hutch++ method.   Our approach is based on employing a randomized low-rank approximation followed by a residual calculation, resulting in statistical errors that scale much better than $\sqrt{N_{\xi}^{-1}}$. We implement the approach within the second-order Born approximation for the self-energy in the computation of neutral excitations and discuss three different low-rank approximations for the two-body Coulomb integrals. Tests on a series of hydrogen dimer chains with varying lengths demonstrate that the Hutch++-like approximations are computationally more efficient than both deterministic and purely stochastic (Hutchinson) approaches for low error thresholds and intermediate system sizes. Notably, for arbitrarily large systems, the Hutchinson-like approximation outperforms both deterministic and Hutch++-like methods.
\end{abstract}

\maketitle

\section{\label{sec:intro}Introduction}
Studying the electronic structure of molecular systems and materials is essential for understanding, predicting, and controlling their properties. An especially intriguing aspect involves the calculation of excited electronic states, as they play a crucial role in photo-chemical transformations,\cite{vlvcek2010ultrafast,hammond1963organic} energy storage and transfer,\cite{hsu2009electronic,welin2017photosensitized,mejia2019effect} and light harvesting.\cite{konig2012quantum} While exact many-body techniques like full configuration interaction (FCI) and tensor network-based methods offer accurate results for simplified problems, they are constrained by computational costs limiting their use to small systems. In the realm of extended systems, approximations such as mean-field time-dependent density functional theory (TDDFT)~\cite{marques2004time,burke2005time,casida2012progress,marques2006time2} or time-dependent Hartree-Fock (TDHF),\cite{mclachlan1964time,jorgensen1975molecular,li2005time}  as well as methods like coupled cluster within the equations of motion formalism (EOM-CC),\cite{hirata2004higher, bartlett2012coupled} are commonly used for calculating excited states. Additionally, techniques based on Green's function (GF) methods, such as GW~\cite{hybertsen1986electron, aryasetiawan1998gw, van2006quasiparticle, kotani2007quasiparticle, golze2019gw, shishkin2007self} and GF2~\cite{phillips2014communication, pavovsevic2017communication} closures, have proven effective for computing excited states in both extended materials and molecular systems.

Within the family of GF techniques, we will explore a method known as GF2 or the second-order Born approximation that approximates electronic correlations by employing a second-order expansion of the self-energy with bare Coulomb interactions.\cite{mejia2023stochastic,doutime2022} Previous evaluations of the accuracy of GF2 have shown favorable comparisons with other methods such as configuration interaction with singles and perturbative doubles (CIS(d)), demonstrating notably precise results for excited states, even for those with charge transfer character.\cite{doutime2022} However, one drawback of the GF2 method is its computationally intensive nature compared to mean-field-based techniques, with a complexity of $O(N^5)$, where $N$ represents the number of basis functions used to describe the system. To address this steep scaling, in Ref.~\onlinecite{mejia2023stochastic}, we utilized the stochastic resolution of the identity (sRI)~\cite{takeshita2017stochastic, dou2020range} to decouple the 4-index electron repulsion integrals (ERI) present in the self-energy. This led to a stochastic real-time implementation of GF2, referred to as stochastic TD-GF2 (or sTD-GF2), which offers a computational scaling of $O(N^3)$ for computing excited states, similar to a mean-field calculation.

The enhanced computational efficiency achieved through the utilization of stochastic techniques is accompanied by the introduction of a statistical error, which can be managed by increasing the number of samples or stochastic orbitals. Particularly, when employing the stochastic resolution of the identity to decouple the ERIs as in sTD-GF2, the convergence of the statistical error is relatively slow, scaling as $O(N_\xi^{-1/2})$, where $N_\xi$ represents the number of stochastic orbitals utilized to approximate the sRI. This convergence behavior resembles that of stochastic trace estimators based on the Hutchinson algorithm.\cite{hutchinson1989stochastic,girard1987algorithme} For the latter, enhancements can be achieved by combining low-rank approximations to the matrix whose trace is calculated with the stochastic estimation of the residual. This approach, termed Hutch++,\cite{meyer2021hutch++} has demonstrated significant success in enhancing error convergence in the stochastic estimation of the trace of positive semi-definite matrices, particularly when they exhibit a mild low-rank structure.

In this study, we present a method akin to Hutch++ for the stochastic resolution of identity and investigate its efficacy in comparison to the traditional Hutchinson-like approach for excited states of hydrogen dimer chains. Drawing inspiration from Hutch++, our proposed method entails decomposing the computation into a low-rank component and a residual. To achieve this, we employ a randomized singular value decomposition (SVD) technique on the 4-index Electron Repulsion Integrals (ERI) and subsequently select stochastic orbitals to approximate the residuals. In Section~\ref{sec:theory}, we offer an overview of the stochastic TD-GF2 method, elaborating on its theoretical underpinnings, and extend the Hutch++ approximation for the resolution of identity. In Section~\ref{sec:results}, we delve into various strategies for constructing a Hutch++-like approximation to the sRI and rigorously evaluate the performance against deterministic TD-GF2 and the Hutchinson-like approach. Finally, in Section~\ref{sec:conclusions}, we provide a summary of our findings, discussing the implications and potential avenues for future research.


\section{\label{sec:theory}Theory}
This section provides an overview of the real-time second-order Green's function theory for computing neutral excitations (TD-GF2) and its stochastic implementation (sTD-GF2), along with the presentation of Hutch++-like variants of the resolution of identity. For a detailed derivation of TD-GF2 and sTD-GF2, readers are referred to Ref.~\onlinecite{doutime2022} and \onlinecite{mejia2023stochastic}, respectively.

\subsection{Deterministic Real-Time GF2 theory (TD-GF2)}
In the TD-GF2 method, we consider a generic many-body Hamiltonian coupled to an external electric field, expressed in second quantization as:
\begin{equation}
    \hat H = \sum_{ij} h_{ij}   \hat a^\dagger_i \hat a_j + \frac12 \sum_{ijkl}(ij|kl)  \hat a^\dagger_i \hat a^\dagger_k  \hat a_l \hat a_j + \sum_{ij} \Delta_{ij}(t) \hat a^\dagger_i \hat a_j~,
    \label{eq:H}
\end{equation}
where $i$, $j$, $k$, and $l$ denote molecular orbital indexes, $\hat{a}^\dagger_i$ and $\hat{a}_i$ represent creation and annihilation operators for an electron in orbital $\chi_i$, respectively, $h_{ij}$ are the matrix elements of the one-body interactions, $(ij|kl)$ are the matrix elements of the two-body interactions, corresponding to 4-index electron repulsion integrals:
\begin{equation} 
 \label{eq:2e4c}
 (ij|kl) = \iint  \frac{\chi_i ({\bf r}_1)\chi_j ({\bf r}_1)\chi_k ({\bf r}_2)\chi_l ({\bf r}_2)}{\left|{\bf r}_1-{\bf r}_2\right|} d{\bf r}_1 d{\bf r}_2.
\end{equation}
The last term in Eq.~\eqref{eq:H} represents the external driving force, $\Delta_{ij}(t)=E(t) \cdot {\bf \mu}_{ij}$, where $E(t)$ is a time-dependent perturbation that couples the system with an external electric field, through its transition dipole moment ${\bf \mu}_{ij}$ defined by:
\begin{equation}
 {\bf \mu}_{ij}=\int \chi_i({\bf r}) {\bf r} \chi_j({\bf r}) d{\bf r}.
 \label{eq:mu_ij}
\end{equation}
Under the adiabatic approximation, where the system responds instantaneously to external stimuli, the equation of motion for the electronic density matrix $\rho(t)$ in the basis of the eigenstates of the Fock operator is given by:\cite{doutime2022,mejia2023stochastic}
\begin{equation}
\label{eq:eom_rho}
i\frac{d}{dt}\rho(t) = [{\bf F}[\rho(t)],\rho(t)] + \tilde{\bf \Sigma}^{ad}(t)\rho(t) - \rho(t)\tilde{\bf \Sigma}^{ad\dag}(t)~,
\end{equation}
where ${\bf F}[\rho(t)]$ is the Fock operator, with matrix elements given by:
\begin{equation}
    F_{ij}[\rho(t)] = h_{ij} + \sum_{kl}(ij|kl)\rho_{kl}(t) -\frac{1}{2} \sum_{kl}(ik|jl)\rho_{kl}(t) + \Delta_{ij}(t)~,
\end{equation}
and $\tilde{\bf \Sigma}^{ad}$ is the adiabatic GF2 self-energy, with matrix elements given by:
\begin{equation}
\label{eq:self_det}
    \tilde{\Sigma}^{\rm ad}_{ij}(t) = -\sum_{mn}\delta\tilde{W}^{R}_{imjn}\rho_{mn}(t)
    +\frac{1}{2}\Re\sum_{mn}\delta\tilde{W}^{R}_{imjn}\delta_{mn}~.
\end{equation}
In the above equation, $\delta\tilde{W}^{R}_{imjn}$ are matrix elements of the GF2 retarded screened Coulomb interactions taken in the zero-frequency limit:
\begin{equation}
\label{eq:W}
\begin{split}
    \delta\tilde{W}^{R}_{imjn} =& \lim_{\omega \to 0}\Bigg\{-\frac{1}{2}\sum_{kq} \frac{f(\varepsilon_k)-f(\varepsilon_q)} {\varepsilon_k-\omega-\varepsilon_q-i\eta}\\
    &\times (im|qk)\Big[2(jn|qk)-(jk|qn)\Big]\Bigg\}~.
\end{split}
\end{equation}
In the equation above, $f(\varepsilon)$ is the Fermi function and $\{\varepsilon\}$ are GF2 quasiparticle energies obtained as in Ref.~\onlinecite{dou2019stochastic}. The dynamics of $\rho(t)$, obtained by integrating Eq.~\eqref{eq:eom_rho}, is used to compute the frequency-dependent absorption spectrum (photoabsorption cross-section) $\sigma(\omega)$ of the system as:
\begin{equation}
    \sigma(\omega) \propto \frac{1}{3\gamma}\sum_{d=x,y,z}\omega\Im\int dt e^{-i\omega t}\rm{Tr}[(\rho(t)-\rho(t_0))\mu_d]~.
\label{eq:abs}
\end{equation}
In the above equation, $d=x,y,z$ are the spatial components of the dipole moment and $\gamma\ll 1$ is a small dimensionless parameter that scales the amplitude of the external electric field. 
The computational cost of the TD-GF2 method is primarily determined by the expense of evaluating the self-energy (Eq.~\eqref{eq:self_det}), which scales as $O(N^5)$ with the system size.

\subsection{Stochastic Real-Time GF2 theory (sTD-GF2)}
To reduce the scaling of the TD-GF2 method from $O(N^5)$ to $O(N^3)$,  the 4-index ERI appearing in the self-energy (Eq.~\eqref{eq:self_det}) are decoupled using the stochastic resolution of the identity (sRI), which corresponds to the average of the outer product of $N_\xi$ stochastic orbitals $\theta$, defined as vectors with random elements $\pm 1$, in the limit of infinite stochastic orbitals:
\begin{equation}
\label{eq:I}
    \lim_{N_\xi\to\infty}
    \mean{ \theta^\xi\otimes{\theta^\xi}^T}_{\xi} =
     \begin{pmatrix}
    1 & 0 & \cdots & 0\\
    0 & 1 & \cdots & 0\\
    \vdots & \vdots & \ddots & \vdots\\
    0 & 0 & \cdots & 1
    \end{pmatrix}
    = {\bf I}~,
\end{equation}
where $\left\langle\cdots\right\rangle_{\xi}$ represents the average over the set $\{\theta^\xi\}$ of uncorrelated stochastic orbitals $\theta^\xi$, with $\xi=1,2,\dots,N_\xi$. 
Using the sRI, the 4-index ERI can be approximately written as~\cite{takeshita2017stochastic}
\begin{equation}
\label{eq:ERI_sRI}
\begin{split}
    (ij|kl)
    \approx&\sum_{PQ}^{N_\text{aux}}\sum_{AB}^{N_\text{aux}} (ij|A)V_{AP}^{-1/2} \left\langle \theta_{P}^\xi \otimes {\theta_{Q}^\xi}^T \right\rangle_{\xi}
     V_{QB}^{-1/2}(B|kl)\\
    &= \left\langle R_{ij}R_{kl} \right\rangle_{\xi}.
\end{split}
\end{equation}
We use capital indexes A and B to refer to the auxiliary basis with $N_{aux}$ elements. In Eq.~\eqref{eq:ERI_sRI}, $R_{ij}=\sum_{A}^{N_{aux}}(ij|A)  \sum_{B}^{N_{aux}}V_{AB}^{-1/2}\theta_B$, and $(ij|A)$ and $V_{AB}$ are 3-index and 2-index ERI, given by
\begin{equation}
    (ij|A) = \int \frac{\chi_i({\bf r}_1)\chi_j({\bf r}_1)\chi_A({\bf r}_2)}{|{\bf r}_1 - {\bf r}_2|}d{\bf r}_1d{\bf r}_2
\end{equation}
and
\begin{equation}
\label{eq:V}
    V_{AB} = \int \frac{\chi_A({\bf r}_1)\chi_B({\bf r}_2)}{|{\bf r}_1 - {\bf r}_2|}d{\bf r}_1d{\bf r}_2~,
\end{equation}
respectively.

The sRI can be seen as applying the Hutchinson method for estimating the trace of a certain matrix (defined for each ERI). Specifically, for a matrix with elements defined as $D_{PQ}=\sum_{AB} (ij|A)V_{AP}^{-1/2} V_{QB}^{-1/2}(B|kl)$, the Hutchinson estimator of its trace is $\left\langle R_{ij} R_{kl}\right\rangle_{\xi}$.

To obtain the self-energy, we insert the 4-index ERI given by Eq.~\eqref{eq:ERI_sRI} into Eq.~\eqref{eq:W} and use it to approximate the self-energy in Eq.~\eqref{eq:self_det}. This yields the following expression for the self-energy:
\begin{equation}
\label{eq:self_sRI}
\begin{split}
    \Sigma^{\rm ad}_{ij}(t) \approx & -\frac{1}{2}\Bigg\langle\sum_{kqmn} \frac{f(E_k)-f(E_q)}{E_k-\omega-E_q-i\eta}\\
    &R_{im}R_{qk}(2R'_{jn}R'_{qk}-R'_{jk}R'_{qn})\delta\rho_{mn}(t)\Bigg\rangle_{{\xi},{\xi'}},
\end{split}
\end{equation}
where the prime symbols over the $R$ tensors indicate that different sets of stochastic orbitals are employed to avoid correlations. When using Eq.~\eqref{eq:self_sRI} to approximate the self-energy, we will refer to the method as the Hutchinson-like approximation of TD-GF2.

Note that the scaling of the self-energy in the Hutchinson-like approximation (Eq.~\eqref{eq:self_sRI}) is $O(N^2_\xi N^3)$. However, we have observed that the error per electron 
in many quantities is independent of
the system size, leading to an effective $O(N^3)$ scaling\cite{mejia2023stochastic} (this is further tested in Sec. \ref{sec:results}). Such system size independent behavior of $N_\xi$ for a given error has been observed in stochastic computations of the ground state energy per electron\cite{takeshita2017stochastic, takeshita2019stochastic, dou2020range}, charged excitation energies\cite{dou2019stochastic} in molecules and nanostructures, and the photoabsorption cross-section per electron, which we study further below.

\subsection{Hutch++-like approximations for TD-GF2}
Hutch++~\cite{meyer2021hutch++} is a method for estimating the trace of positive semidefinite matrices with mild low-rank structure. The method consists of building a stochastic low-rank approximation of the matrix by performing a randomized SVD with a test matrix ${\bf W}$ (consisting of $N_\xi$ stochastic vectors as columns) as it is shown in Algorithm~\ref{al:svd}. Then, a deterministic trace is performed on the low-rank ${\bf A}^\text{low-rank}$, while the residual is estimated stochastically (Hutchinson algorithm~\cite{hutchinson1989stochastic, girard1987algorithme}) using the residual stochastic vectors, corresponding to the columns of ${\bf G}$ (see Algorith~\ref{al:svd}). The advantage of the Hutch++ algorithm over the direct use of the sRI (Hutchinson algorithm) is that the convergence of the statistical error is faster than $O(\sqrt{N_\xi^{-1}})$.

\begin{figure}[H]
\begin{algorithm}[H]
  \caption{Randomized low-rank of ${\bf A}$}
  \label{al:svd}
   \begin{algorithmic}[1]
   \State ${\bf W} \equiv [\theta^{1}\cdots\theta^\xi]$ Matrix with stochastic column orbitals
   \State ${\bf Q} = \mathrm{orth}[{\bf AW}]$ Orthogonalize using QR decomposition
    \State ${\bf A}^\text{low-rank}={\bf QQ}^T{\bf A}$
   \State ${\bf G} = ({\bf I} - {\bf Q}{\bf Q}^T) {\bf W}' \equiv [G^{1'}\cdots G^{\xi'}]$ Matrix with residual stochastic column orbitals
   \end{algorithmic}
\end{algorithm}
\end{figure}

 \begin{figure*}[t]
 \centering
\includegraphics[scale=1.0]{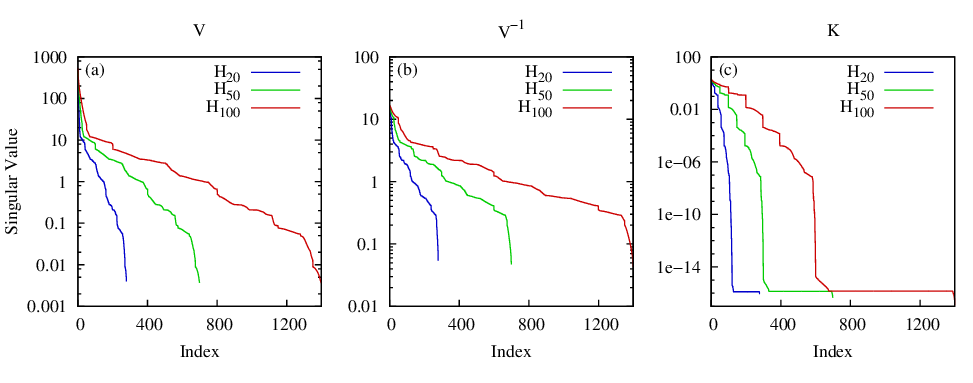}
\caption{Singular values of (a) the Coulomb matrix ${\bf V}$, (b) its inverse ${\bf V}^{-1}$, and (c) the reshaped 3-index K-tensor ${\bf K}$ for three representative hydrogen dimer chains. The Coulomb matrix and its inverse were computed using the cc-PVDZ-RI auxiliary basis set, while the K-tensor utilized the minimal STO-3G basis as the main basis and cc-PVDZ-RI as the auxiliary basis.}
\label{fig:singular}
\end{figure*}

Following the logic of the Hutch++ method to accelerate the convergence of stochastic trace estimations, we have developed three Hutch++-like variants of the sTD-GF2 approach. As before, we rewrite the 4-index ERI in terms of 3-index and 2-index ERIs:
\begin{equation}
    (ij|kl) \approx \sum_{AB}^{N_\text{aux}}(ij|A)V_{AB}^{-1}(B|kl)~.
\end{equation}
However, instead of using the stochastic resolution of identity given by Eq.~\eqref{eq:I}, we first perform a randomized low-rank approximation to  ${\bf V}$, ${\bf V}^{-1}$, or   $K_{ij}^Q=\sum_A^{N_\text{aux}}(ij|A)V_{AQ}^{-1/2}$, and compute the residuals stochastically. The latter is typically used to approximate the ERIs deterministically (this approximation is called density fitting) as $(ij|kl)=\sum_Q^{N_\text{aux}}K_{ij}^Q K_{kl}^Q$. We will refer to these Hutch++-like variants of sTD-GF2 as H++ on $V$, H++ on $V^{-1}$, and H++ on $K$, respectively. 

Before we describe the different low-rank approximations, we provide in Fig.~\ref{fig:singular} analysis of the singular values of the Coulomb matrix ${\bf V}$, its inverse ${\bf V}^{-1}$, and the reshaped tensor ${\bf K}$ for hydrogen dimer chains of varying lengths (see below in Sec.~\ref{sec:results} more detail on the calculations). In all instances, the rank grows like $O(N)$, indicating that any truncation based on low-rank approximation will inherently vary with system size. This variability might result in an increase in the method's scaling with system size, compared to the purely stochastic approach (such as the Hutchinson-like sTD-GF2). Nevertheless, the reduction in computational overhead due to accelerated error convergence might offset the scaling increase within a specific parameter range. This aspect will be investigated numerically in Sec. \ref{sec:results}.\\

\subsubsection{\label{sec:H++V} Low rank approximations based on ${\bf V}$ and ${\bf V}^{-1}$}
For the H++ on $V$ and H++ on $V^{-1}$ variants, the 4-index ERI can be expressed as a sum of low-rank and residual terms:
\begin{equation}
\label{eq:ERI_hutch}
\begin{split}
    (ij|kl) \approx \sum_{r}^{N_\text{rank}}N_{ij}^r M_{kl}^r + \left\langle R_{ij}^{\xi'}R_{kl}^{\xi'} \right\rangle_{\xi'}~,
\end{split}
\end{equation}
where the tensors elements $N_{ij}^r$, $M_{kl}^r$, and $R_{ij}^{\xi'}$ are defined in Appendix A. The first term in Eq.~\eqref{eq:ERI_hutch} is a randomized low-rank approximation of the ERI and the second term corresponds to the residual, which is estimated stochastically.

The scaling of computing the $N_{ij}^r$ and $M_{kl}^r$ tensors is $O(N_\text{rank}N^3)$, where $N_\text{rank}$ is the rank of the stochastic low-rank approximation, which coincides with the number of stochastic orbitals used to build the test matrix ${\bf W}$ (see Algorithm \ref{al:svd}), i.e. $N_\text{rank}=N_\xi$.

Using Eq.~\eqref{eq:ERI_hutch} to evaluate the ERIs, the self-energy given by Eq.~\eqref{eq:self_det}, can be expressed as:
\begin{widetext}
    \begin{equation}
    \label{eq:self_Hutch}
        \begin{split}
             \delta\Sigma^{\rm ad}_{ij}(t) =& -\frac{1}{2}\sum_{r,r'}^{N_\text{rank}}\sum_{kqmn} \frac{f(\varepsilon_k)-f(\varepsilon_q)}{\varepsilon_k-\omega-\varepsilon_q-i\eta}N_{im}^{r}M_{qk}^{r}(2N_{jn}^{r'}M_{qk}^{r'} -N_{jk}^{r'}M_{qn}^{r'})\delta\rho_{mn}(t)\\
             & -\frac{1}{2}\Bigg\langle\sum_{r}^{N_\text{rank}}\sum_{kqmn} \frac{f(\varepsilon_k)-f(\varepsilon_q)}{\varepsilon_k-\omega-\varepsilon_q-i\eta}N_{im}^{r}M_{qk}^{r}(2R'_{jn}R'_{qk} -R'_{jk}R'_{qn})\delta\rho_{mn}(t)\Bigg\rangle_{\xi'}\\
             & -\frac{1}{2}\Bigg\langle\sum_{r'}^{N_\text{rank}}\sum_{kqmn} \frac{f(\varepsilon_k)-f(\varepsilon_q)}{\varepsilon_k-\omega-\varepsilon_q-i\eta}R_{im}R_{qk}(2N_{jn}^{r'}M_{qk}^{r'} -N_{jk}^{r'}M_{qn}^{r'})\delta\rho_{mn}(t)\Bigg\rangle_{\xi}\\
             & -\frac{1}{2}\Bigg\langle\sum_{kqmn} \frac{f(\varepsilon_k)-f(\varepsilon_q)}{\varepsilon_k-\omega-\varepsilon_q-i\eta}R_{im}R_{qk}(2R'_{jn}R'_{qk} -R'_{jk}R'_{qn})\delta\rho_{mn}(t)\Bigg\rangle_{\xi,\xi'}~.
        \end{split}
    \end{equation}
\end{widetext}
The first term on the right-hand side of Eq.~\eqref{eq:self_Hutch} corresponds to the randomized low-rank approximation of the self-energy, with a scaling of $O(N_\text{rank}^2 N^3)$. The next two terms are mixed stochastic and low-rank, with a scaling of $O(N_\text{rank} N^3)$. The last term in Eq.~\eqref{eq:self_Hutch} is purely stochastic, scaling as $O(N^3)$. As shown above in Fig.~\ref{fig:singular}, the low-rank structure of the matrices becomes less pronounced for larger systems, implying that $N_\text{rank}$ does depend on $N$. However, we expect $N_\text{rank} < N$, so $O(N^3) < O(N_\text{rank}^2 N^3) < O(N^5)$, corresponding to the scaling of the Hutchinson-like, Hutch++-like, and deterministic TD-GF2 approximations, respectively.

\begin{figure}[h!]
\centering
\includegraphics[scale=1.0]{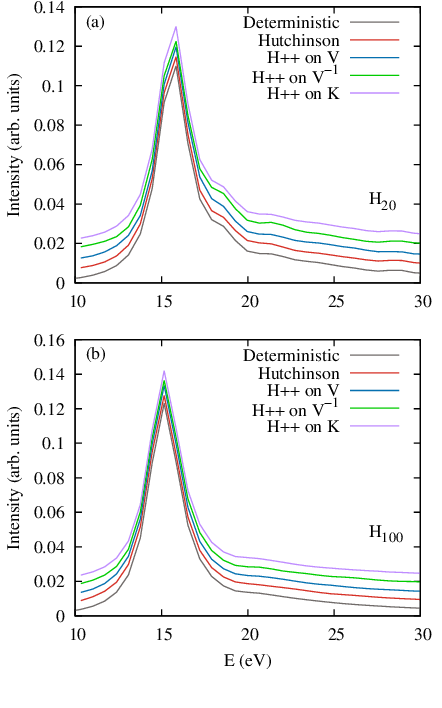}
\caption{Absorption spectra of representative hydrogen dimer chains, with (a) H$_{20}$ and (b) H$_{100}$, were analyzed using deterministic (TD-GF2), Hutchinson-stochastic (sDT-GF2), and mixed low-rank Hutch++-like variations of sTD-GF2 applied to the Coulomb matrix (H++ on ${\bf V}$), its inverse (H++ on ${\bf V}^{-1}$), and the 3-index tensor $K$ (H++ on ${\bf K}$). In all cases, the STO-3G basis set was utilized as the primary basis, with cc-PVDZ-RI employed as the auxiliary basis. The intensity average error per electron was set to $10^{-3}$, with units consistent with the intensity. The curves have been vertically shifted for clarity}
\label{fig:spectra}
\end{figure}

\begin{figure*}[t]
\centering
\includegraphics[scale=1.0]{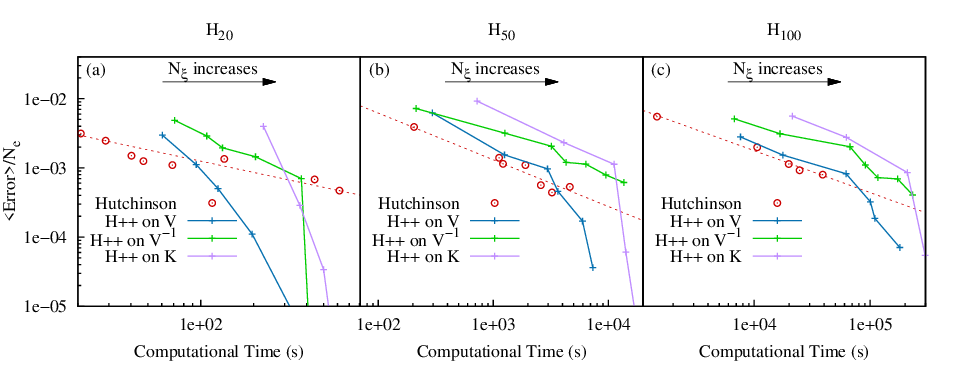}
\caption{Error per electron in computing the absorption spectra of representative hydrogen dimer chains, namely (a) H$_{20}$, (b) H$_{50}$, and (c) H$_{100}$, as a function of computational effort (time) using various stochastic methods. Computational time increases with decreasing error as a result of the increase in the number of stochastic orbitals, $N_\xi$. In all cases, the STO-3G basis set was utilized as the main basis, and cc-PVDZ-RI was employed as the auxiliary basis.}
\label{fig:error}
\end{figure*}

\begin{figure}[t]
 \centering
\includegraphics[scale=1.0]{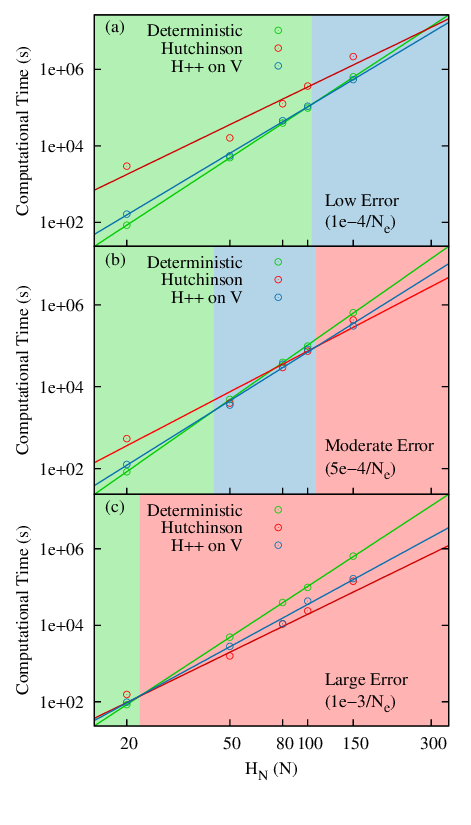}
\caption{Computational scaling with system size for the deterministic (TD-GF2), Hutchinson-stochastic (sTD-GF2), and mixed low-rank sTD-GF2-Hutch++ methods applied to the Coulomb matrix (H++ on ${\bf V}$) for (a) low ($10^{-4}$), (b) moderate ($\frac{1}{2} \times 10^{-3}$), and (c) large ($10^{-3}$) error per electron across a series of hydrogen dimer chains. The green, blue, and red shaded regions indicate the range of system sizes where TD-GF2, sTD-GF2, and sTD-GF2-Hutch++ (H++ on ${\bf V}$) are most efficient, respectively.}
\label{fig:scaling}
\end{figure}

\subsubsection{\label{sec:H++K} Low rank approximation based on K}
The ERIs using the H++ on $K$ variant are given by
\begin{equation}
\label{eq:ERI_K}
\begin{split}
    (ij|kl) =& \sum_r^{rank}N_{ij}^r M_{kl}^r + \mean{(R_{ij}^\xi)^\text{low-rank}R_{kl}^\xi}_\xi\\
    &+ \mean{R_{ij}^\xi (R_{kl}^\xi)^\text{low-rank}}_\xi + \mean{R_{ij}^\xi R_{kl}^\xi}_\xi~,
\end{split}
\end{equation}
where all tensor elements are defined in Appendix A. The first term in Eq.~\eqref{eq:ERI_K} is a randomized low-rank approximation of the ERI, the second and third terms are mixed stochastic and low-rank, and the last term corresponds to the stochastic residual. Using Eq.~\eqref{eq:ERI_K} to approximate the ERIs in the self-energy given by Eq.~\eqref{eq:self_det} results in the $H$++ on $K$ variant of sTD-GF2.

Despite the common ingredients with Hutch++, all of the sTD-GF2 variants proposed here (H++ on ${\bf V}$, H++ on ${\bf V}^{-1}$, and H++ on ${\bf V}$) are substantially different from the Hutch++ trace estimator since using the deterministic subspace for computing the low-rank approximations changes the scaling of the method. Therefore, the implications for their performance must be tested. In the next section, we numerically analyze the efficiency of the proposed schemes.

\section{\label{sec:results}Results}
To test the performance of the Hutch++-like variants of sTD-GF2, we consider hydrogen dimer chains of varying lengths as model systems. The hydrogen chains, consisting of hydrogen molecules with a bond length of 0.74 \AA~ and inter-molecular distance of 1.26 \AA, were aligned along the direction of the external electric field. To represent the electric field, we employed a Gaussian pulse centered at $t_0=0.2$ fs with an amplitude $\gamma E_0 = 0.02$ V/\AA~ and variance of 0.005 fs. The regularization parameter in Eq.\eqref{eq:W} was set to $\eta=0.01$ and the inverse temperature to $\beta=50$. To perform the Fourier transform in Eq.\eqref{eq:abs}, we have added the damping function $e^{-\Gamma t}$, with a decay rate $\Gamma=1/(0.1t_\text{max})$, being $t_\text{max}$ the total propagation time. Throughout this work, we used the minimal STO-3G basis set as the main basis and the cc-PVDZ-RI as the auxiliary basis for decoupling the ERIs. 

Firstly, we aim to illustrate the congruence between the absorption spectra as described by Eq.~\eqref{eq:abs}, calculated through the Hutchinson-like (sTD-GF2) and Hutch++-like variants of sTD-GF2, with the deterministic outcome (TD-GF2). To achieve this, we constrain the dynamics generating $\sigma(\omega)$ to a duration of $6$~fs, resulting in a relatively broad spectral characteristic. Fig.~\ref{fig:spectra} exhibits the absorption spectra for two hydrogen dimer chains of varying lengths, chosen as representative examples. Within the stochastic outcomes' margin of noise, all spectra demonstrate a notable consistency with the deterministic TD-GF2 outcome. Note that the curves have been shifted for clarity.

Next, we shift our focus to examining the statistical error and evaluating its correlation with the number of stochastic orbitals and computational time. Fig.~\ref{fig:error} presents the average statistical error per hydrogen dimer across three typical chains of differing lengths, employing the Hutchinson sTD-GF2 method and its Hutch++-like variations over varying computational durations. The growth in computational time within a specific method arises from increasing $N_\xi$.
The error metric was determined as the mean of the standard error --defined as the ratio between the standard deviation (STD) and the number of independent stochastic runs ($n$)-- of the photoabsorption cross-section:
\begin{equation}
\label{eq:error}
    {\left\langle \mathrm{Error} \right\rangle} = \left\langle \frac{\mathrm{STD}[\sigma(\omega)]}{\sqrt{n}} \right\rangle_{\omega}~,
\end{equation}
where, $\langle\cdots\rangle_\omega=\frac{1}{N_\omega}\sum_\omega^{N_\omega}\cdots$ is the average over the $N_\omega$ discretized frequencies,
$
    \mathrm{STD}[\sigma(\omega)] = \sqrt{\sum_i^n(\sigma_i(\omega)-\mean{\sigma(\omega)})^2/n}
$, and $i=1,\dots,n$ is the index that labels the independent stochastic runs, taken to be $n=6$. We confine our analysis to the scientifically significant interval of $10-30$~eV, as this is the range in which significant absorption peaks are observed for hydrogen dimer chains (see Fig. \ref{fig:spectra}).

We observe that across all instances, the convergence rate of the statistical error remains roughly consistent with that of the standard Hutchinson method when utilizing a limited number of stochastic orbitals. However, it accelerates as the contributions from the low-rank component become more pronounced due to the escalation of $N_\xi$ (remembering that $N_\xi$ governs the rank of this component). Note that, although the singular values of the ${\bf K}$ tensor exhibit the fastest decay among the matrices analyzed in Fig. \ref{fig:singular}, the overhead added by the 16 terms required to compute the square of the ERIs (see Eq.~\eqref{eq:ERI_K}) appearing in the self-energy (Eq.~\eqref{eq:self_det}-\eqref{eq:W}) make the H++ on ${\bf K}$ slower than H++ on ${\bf V}$ and H++ on ${\bf V}^{-1}$, for the range of parameters considered in Fig. \ref{fig:error}. Notably, among all Hutch++ variations, the most rapid convergence occurs when employing randomized Singular Value Decomposition (SVD) on the Coulomb matrix ${\bf V}$. Consequently, we will exclusively focus on this variant within the sTD-GF2 method, alongside the standard Hutchinson and deterministic approaches, for the remainder of the analysis.

Fig.~\ref{fig:scaling} illustrates the computational scaling with the system size for the deterministic TD-GF2, Hutchinson sTD-GF2, and the H++ variant applied to ${\bf V}$ within sTD-GF2 across low, moderate, and high error thresholds. The deterministic approach scales formally as $O(N^5)$, the Hutchinson sDT-GF2 scales as $O(N^3)$ for a fixed error, and the Hutch++ sTD-GF2 as $O(N_\text{rank}^2N^3)$ for a fixed error. Empirically it scales as $O(N^4)$ within the system size range examined in Fig.~\ref{fig:scaling}. This is due to the dependence of $N_\text{rank}$ on $N$. These computational scalings result in a crossover between the preferred method, contingent upon the specified error threshold, and the system size. Specifically, as depicted in Fig.~\ref{fig:scaling}a, the H++ variant on ${\bf V}$ within sTD-GF2 demonstrates superior efficiency compared to the Hutchinson version for systems containing up to approximately $\sim$350 electrons for low error threshold ($1\times 10^{-4}/N_e$ in the error per electron). Conversely, as demonstrated in Fig.~\ref{fig:scaling}c, under a high error threshold ($1\times 10^{-3}/N_e$), the Hutchinson algorithm outperforms the H++ variant on ${\bf V}$ for a broader spectrum of system sizes ($N_H>20$). Fig.~\ref{fig:scaling}b depicts a scenario at intermediate error thresholds.

Additionally, it is worth noting that as error thresholds decrease, both the Hutchinson and H++ variants applied to ${\bf V}$ within sTD-GF2 begin to show disadvantages compared to purely deterministic approach (TD-GF2). This underscores the nontrivial nature of balancing error thresholds and system size when determining the optimal parameter range for the most efficient utilization of the H++ variant on ${\bf V}$. In Fig.~\ref{fig:scaling}, the background color signifies the combination of error threshold and system size where each real-time GF2 method (green for TD-GF2, red for sTD-GF2, and blue for the H++ variant on ${\bf V}$ within sTD-GF2) exhibits the highest computational efficiency. 

Broadly speaking, the H++ variant applied to ${\bf V}$ within sTD-GF2 proves to be more efficient than both purely deterministic and purely stochastic computations for intermediate system sizes and low to moderate error thresholds. For instance, this efficiency is evident within the range of approximately $\sim 100-350$ electrons (100-350 STO-3G basis functions and 1400-4900 cc-PVDZ-RI basis functions) under an error threshold of $10^{-4}$ per electron and $\sim 48-110$ electrons (48-110 STO-3G basis functions and 672-1540 cc-PVDZ-RI basis functions) under an error threshold of $\frac{1}{2} \times 10^{-3}$ per electron, as calculated according to Eq.~\ref{eq:error}.

\section{\label{sec:conclusions}Conclusions}

We have devised a series of Hutch++-inspired iterations of the stochastic real-time GF2 approach tailored for computing neutral excitations. The Hutch++ methodology involves breaking down the computation of the self-energy and the ERIs into a randomized low-rank segment and a stochastically estimated residual. Our work has demonstrated that employing a randomized Singular Value Decomposition technique on the Coulomb matrix $V$ to obtain the low-rank approximation to the ERIs (H++ on $V$) yields the swiftest balance between statistical error and computational time. Nevertheless, the efficiency of such an approximation diminishes for large systems due to its scaling of $O(N_\text{rank}^2N^3)$. However, this efficiency threshold shifts towards larger system sizes when the predetermined statistical error threshold is set to be small.

Generally, the H++ on $V$ proves to be more efficient than both deterministic and stochastic sTD-GF2 methods for scenarios involving low error thresholds and intermediate system sizes. However, the applicability of this parameter range is limited. Consequently, for large system sizes, the Hutchinson sTD-GF2 approach is poised to outperform both the deterministic and Hutch++ variants, owing to its $O(N^3)$ scaling.

The Hutch++-inspired variations of the sTD-GF2 method serve as valuable additions to the arsenal of noise reduction strategies employed in electronic structure calculations. These variants synergize effectively with other techniques, such as the range-separated sRI,\cite{dou2020range} which takes advantage of the sparsity of the ERIs by partitioning them into distinct categories --large contributions, calculated deterministically, and small contributions, computed stochastically-- This partition provides an alternative avenue for enhancing the rate of error convergence in stochastic applications based on the resolution of identity.

\acknowledgements
ER and GKC are grateful for support from the U.S. Department of Energy, Office of Science, Office of Advanced Scientific Computing Research and Office of Basic Energy Sciences, Scientific Discovery through Advanced Computing (SciDAC) program, under Award No. DE-SC0022088. Some of the methods used in this work were provided by the "Center for Computational Study of Excited State Phenomena in Energy Materials (C2SEPEM)", which is funded by the U.S. Department of Energy, Office of Science, Basic Energy Sciences, Materials Sciences, and Engineering Division, via Contract No. DE-AC02-05CH11231 as part of the Computational Materials Sciences program. Resources of the National Energy Research Scientific Computing Center (NERSC), a U.S. Department of Energy Office of Science User Facility operated under Contract No. DE-AC02-05CH11231 are also acknowledged.


\appendix

\renewcommand{\thefigure}{A\arabic{figure}}
\renewcommand{\thetable}{A\arabic{table}}
\setcounter{figure}{0}
\setcounter{table}{0}
\setlength\extrarowheight{10pt}

\section{Expressions for tensor elements}

\subsection{H++ on $V$ and H++ on $V^{-1}$}

The tensor elements for computing the low-rank terms in Eq.~\eqref{eq:ERI_hutch} are given by
\begin{equation}
\begin{split}
    &\left[N_{ij}^{r}\right]^\text{H++ on $V$} = \sum_{A}^{N_\text{aux}} (ij|A) \left( \sum_P V^{-1}_{AP}Q_{Pr} \right)\\
&\left[N_{ij}^{r}\right]^\text{H++ on $V^{-1}$} = \sum_{A}^{N_\text{aux}} (ij|A) Q_{Ar}~,
\end{split}
\end{equation}
and
\begin{equation}
\begin{split}
   &\left[M^r_{kl}\right]^\text{H++ on $V$} = \sum_B^{N_\text{aux}} Q_{rB}^T (B|kl) \\
&\left[M^r_{kl}\right]^\text{H++ on $V^{-1}$} = \sum_B^{N_\text{aux}}\left( \sum_C^{N_\text{aux}} Q_{rC}^T V_{cB}^{-1}
 \right)(B|kl)~,
\end{split}
\end{equation}
for the H++ on $V$ and H++ on $V^{-1}$ methods. Here, $Q_{Pr}$ is obtained following Algorithm~\ref{al:svd}, with ${\bf A}={\bf V}$ or ${\bf V}^{-1}$. For both methods, the residual tensors in Eq.~\eqref{eq:ERI_hutch} are given by $R_{ij}^{\xi'} = \sum_{A}^{N_\text{aux}} (ij|A) L_A^{\xi'}$ with $L_{A}^{\xi'} = \sum_{P}^{N_\text{aux}} V_{AP}^{-1/2} G_P^{\xi'}$, where $G_P^{\xi'}$ is obtained following Algorithm~\ref{al:svd}.

\subsection{H++ on $K$}

The tensor elements in eq.~\eqref{eq:ERI_K} are given by 
\begin{equation}
    \left[N_{ij}^{r}\right]^\text{H++ on $K$} = Q_{(ij)r}
\end{equation}
 and
\begin{equation}
    \left[M_{kl}^{r}\right]^\text{H++ on $K$} = \sum_{s}^{N_\text{rank}}\sum_{P}^{N_\text{aux}}C_{rP}D_{Ps}Q^T_{s(kl)}~.
\end{equation}
In the above equations, ${\bf Q}$ is obtained as in Algorithm~\ref{al:svd}, for ${\bf A}={\bf K}$ with reshaped elements $K_{mnQ}\to K_{(mn)Q}$, $C_{rP}=\sum_{mn}^N Q_{r(mn)}^T K_{(mn)P}$, and $D_{Ps}=\sum_{mn}^N K_{P(mn)}^\dag Q_{(mn)s}$. The low-rank stochastic tensor, $(R_{ij}^{\xi})^\text{low-rank}$, appearing in Eq.~\eqref{eq:ERI_K},  is given by
\begin{equation}
    (R_{ij}^{\xi})^\text{low-rank} = \sum_{Q}^{N_\text{aux}}K^\text{low-rank}_{(ij)Q} W_Q^{\xi}~,
\end{equation}
with
\begin{equation}
    K^\text{low-rank}_{(ij)Q}=\sum_r^{N_\text{rank}} Q_{(ij)r}\left( \sum_{mn}Q_{r(mn)}^T K_{(mn)Q} \right)~,
\end{equation}
where ${\bf W}$ is the stochastic test matrix, as in Algorithm~\ref{al:svd}. The stochastic tensor, $(R_{ij}^{\xi})$, appearing in the same equation, takes the form:
\begin{equation}
    R_{ij}^{\xi} = \sum_{Q}^{N_\text{aux}}K^\text{res}_{(ij)Q} W_Q^{\xi}~,
\end{equation}
with ${K}_{(ij)Q}^\text{res} = K_{(ij)Q} - K_{(ij)Q}^\text{low-rank}$.

\bibliography{references}
\end{document}